# Heterogeneous antiferroelectric ordering in NaNbO$_3$-SrSnO$_3$ ceramics revealed by direct superstructure imaging

Leonardo Oliveira,[1,*] Mao-Hua Zhang,[2,3] Jurij Koruza,[4] Raquel Rodiquez-Lamas,[5] Can Yildirim,[5] Hugh Simons[1,†]

[1]*Department of Physics, Technical University of Denmark, 2800 Kgs. Lyngby, Denmark*
[2]*Wuzhen Laboratory, Jiaxing, China*
[3]*Department of Materials and Earth Sciences, Technical University of Darmstadt, 64287 Darmstadt, Germany*
[4]*Institute for Chemistry and Technology of Materials, Graz University of Technology, Stremayrgasse 9, 8010 Graz, Austria*
[5]*European Synchrotron Radiation Facility - ESRF, 71 Avenue des Martyrs, CS40220, 38043 Grenoble Cedex 9, France*

Corresponding authors: leosdo@dtu.dk,[*] husimo@fysik.dtu.dk[†]

**Abstract**

NaNbO$_3$-*based* antiferroelectric materials offer a promising pathway towards greener and more cost-effective energy storage devices. However, their intrinsic structural instabilities often lead to reduced energy density that compromises their performance and longevity. In this brief communication, we demonstrate how Dark-Field X-ray Microscopy – when carried out on the characteristically weak 1/4{843}$_{pc}$ superstructure reflection – can map the antiferroelectric phase and its strain heterogeneity in typically small, deeply-embedded grains of a NaNbO$_3$ and 0.95NaNbO$_3$-0.05SrSnO$_3$ ceramics, representative of different phase transition behavior. Our results clearly evidences the stabilizing effect of SrSnO$_3$ on the antiferroelectric phase by enhancing the degree of mesostructural order. In doing so, our method establishes a new platform for exploring the impact of disorder on the long-range strain heterogeneity within antiferroelectrics and other materials with modulated crystal structures.

**Introduction**

Antiferroelectric (AFE) materials are promising candidates for an emerging generation of dielectric capacitors that feature superior energy density for non-linear dielectrics and high power density at high operational frequencies.[1,2]. Such applications demand fast and controlled transitions between the AFE and ferroelectric (FE) states under applied electric fields, which in turn require a deep understanding of how these transitions are affected by the plethora of defects that exist in perovskite oxides. Although it is well known that the structure-property relationship in classical FE materials is highly sensitive to domain sizes and the distribution of local heterogeneities,[3] much less is known when it comes to AFE materials, particularly for lead-free compositions. Therefore, advancing the application of lead-free AFE materials requires novel approaches of characterizing and understanding how local sources of structural disorder may benefit or hinder the macroscopic properties, with an ultimate view to engineering their performance and functionality.

Among the few known lead-free AFE compounds, $NaNbO_3$ (NN) serves as a prime example of how structural heterogeneity affects the AFE-FE transition and, ultimately, its energy storage potential.[4,5] This is particularly due to its complex sequence of temperature-driven phase transitions,[6] grain-size-induced phase transitions,[7] and room temperature polymorphic instability.[5] All these factors pose major challenges on the practical design of AFE NN, by favoring the non-reversible transition to the FE state and thus lowering the recoverable energy density. Various strategies have been proposed for promoting the reversibility of the transition, ranging from ensuring adequate temperature control upon cooling towards room temperature to suppress local shear-strain coupling at the grain level,[8] to alloying NN-*based* solid solutions with other perovskites end members such as $SrSnO_3$, $CaZrO_3$, $BiMg_{2/3}Nb_{1/3}O_3$, and $AgNbO_3$-$CaHfO_3$.[9–12] Some binary systems have shown superior performance, with better-defined double hysteresis loops and recoverable energies and efficiencies higher than 2 J/cm$^3$ and 80%, respectively. However, quantifying and correlating the structural origin of this improvement with the macroscopic performance remains a challenging task that requires high penetration depths to resolve mesostructural features deeply embedded within the bulk, as well as sensitivity across multiple length scales from the macroscopic sample geometry to atomic-scale structural disorder.

Here in this letter, we demonstrate that probing the typically weak superstructure reflections with Dark-Field X-ray Microscopy (DFXM) can be used to directly map the AFE phase, as well as to quantify its structural heterogeneities (*e.g.,* strains) deeply embedded within the bulk of unmodified NN and 0.95NaNbO$_3$-0.05SrSnO$_3$ (abbreviated as NN-0.05SS). These ceramics are representative of the NN-*based* materials with irreversible and reversible field-induced phase transition, respectively. Uniquely, we achieve this by directly mapping the relative elastic strain ($\bar{\varepsilon}_{33}$) from the characteristic 1/4-*type* superstructure reflection associated with the AFE *P*-phase (*Pbcm* space group). Such measurements of superstructure reflections in polycrystalline ceramics are challenging due to the relatively weak scattered signal (typically 2% of the most intense fundamental reflection) and the typically small grain sizes of AFE NN-*based* materials (average grain size < 9 μm). Despite these challenges, using DFXM to quantify spatial lattice deformation associated with superstructure reflection offers several benefits: Firstly, it provides a direct correlation to the stability of the AFE domains within a grain. Secondly, it is a non-destructive technique, and therefore enables direct mapping of the electromechanical boundary conditions while preserving its original residual stress state (to which ferroic materials are notoriously sensitive).[13]

The DFXM experiments were carried out at ESRF-ID06 according to the general measurement protocol described in the literature.[14] Monochromatic X-rays at an energy of 17 keV were used to image the pseudocubic (*pc*) 1/4{843}$_{pc}$ superstructure reflection, with a magnification of 17.4×, and an effective pixel size of 37 × 91 nm$^2$ (*h* and *v*, respectively). Condenser optics were also used to create a line-shaped illumination of approximately 400 × 0.6 μm$^2$ (*h* and *v*, respectively) that ensures that the measured images corresponded to a finite 'slice' within the sample volume with a final voxel size of 37 × 91× 600 nm$^3$ (*h*, *w*, and *d*, respectively). The raw images were acquired by tilting the diffraction vector ($\vec{Q}$) in a range of 0.4°. The grains chosen in our investigation were those representing the most prominent diffracted signal (*i.e.,* in the 2*θ* and azimuthal angle in reciprocal space) within the Debye ring corresponding to the 1/4{843}$_{pc}$. The selected grains were then positioned such that their diffracted intensity was maximized, *i.e.,* at the thickest portion of the diffracting grain. The reconstructed maps of the $\bar{\varepsilon}_{33}$ and FWHM (full-width at half maximum relative to variance of the position superstructure *Q*-vector) were generated by the Gaussian moments approach using the *Darfix* library.[15] The main codes used herein are available at [*link to be added at acceptance*]. Further details on the sample preparation method and dielectric characterization for the ceramic pellets employed in this work are described in the literature.[4]

The DFXM measurements in **Figure 1** show distinct differences in the mesoscale structure between the embedded grains of pure NN and NN-0.05SS. It's important to point out that the shapes depicted in **Figure 1** do not delineate typical grain boundaries, as usually observed in DFXM maps probing fundamental reflections of grains. Instead, they are confined to the intrinsic AFE phase boundary and delineate the volume within the grain where the scattering originates from the superstructure reflection we chose to probe. The integrated intensity of the $1/4\{843\}_{pc}$ reflection shown in **Figure 1(a)** reveals that the AFE phase exists in distinct regions, with those in pure NN (top) being smaller and more numerous compared to NN-0.05SS (bottom).

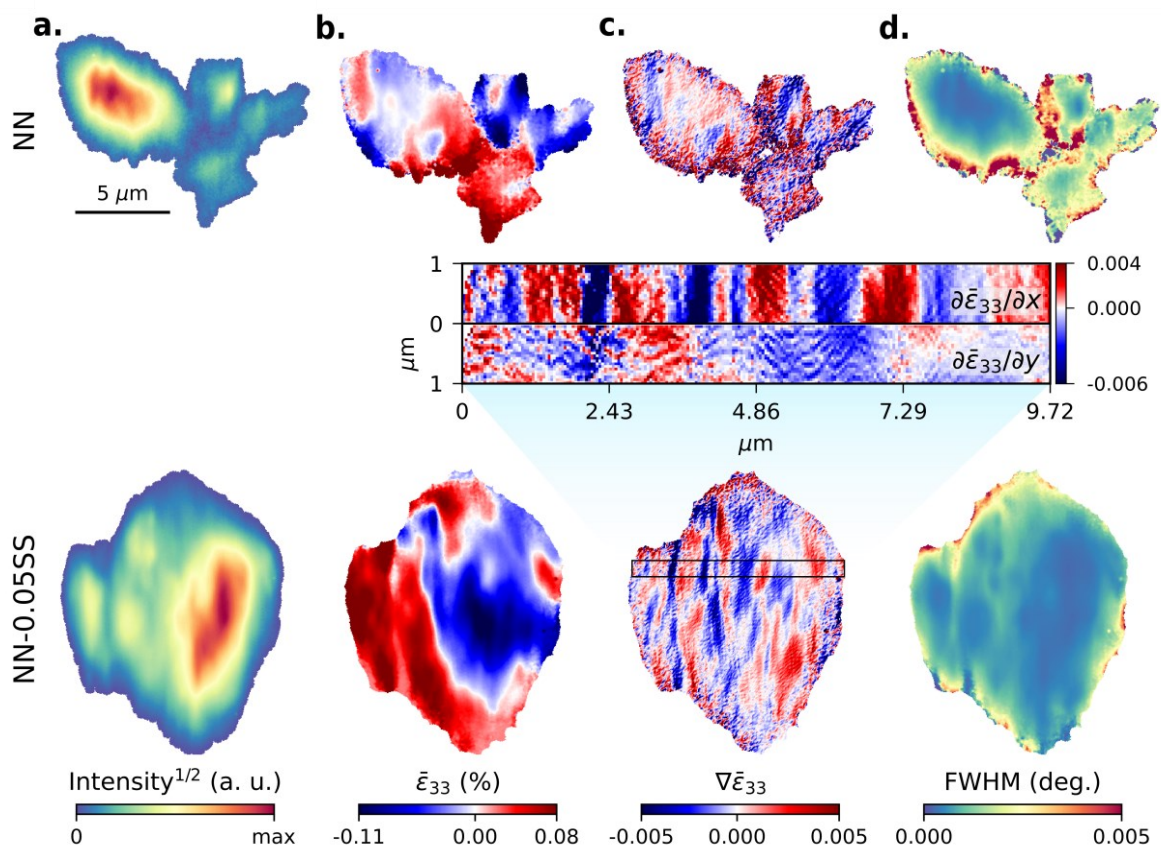

**Figure 1.** Reconstructed DFXM maps of $1/4\{843\}_{pc}$ superstructure refection for NN pure and NN-0.05SS. The boundaries of the probed area are confined to the region where only the AFE phase scatters within the grain. (a) Total scattered intensity, (b) relative elastic strain, (c) strain gradient, and (d) FWHM. The insert in (c) highlights the partial strain gradients and its periodicity along $\partial\bar{\varepsilon}_{33}/\partial x$. The scale bar pertains to all maps displayed here.

In **Figure 1(b)**, the reconstructed $\bar{\varepsilon}_{33}$ maps, *i.e.,* the relative strain along the $1/4\{843\}_{pc}$ zone axis, indicate that these regions can (but not always) be associated with different lattice spacings in a manner consistent with ferroelastic twinning. Remarkably, the relative elastic strain gradient ($\nabla\bar{\varepsilon}_{33}$) of NN-0.05SS, shown in **Figure 1(c)** and inset, depicts clear domains within these regions, typically arranged in chevron-*like* clusters.

Finally, in **Figure 1(d)** the FWHM of the 1/4{843}$_{pc}$ reflection, which scales to the microstrain within the sub-μm voxels, indicates a higher degree of structural disorder both at the outermost AFE phase boundaries (which in parts correlated to grain boundary), as well as the boundaries between the depicted major AFE regions. However, minimal variation is discernible within them, *i.e.,* across the chevron-*like* structures. The DFXM results are therefore consistent with heterogeneous mesostructures in which major ferroelastic twins are separated by regions of high structural disorder and low density of the AFE phase, but that these major regions contain ordered, coherent AFE/ferroelastic domains within them arranged in chevron-*like* patterns with little-to-no reduction in the strength of the AFE structural order.

Meanwhile, the addition of SrSnO$_3$ to NN as a substitution is correlated to an increase in the size of these coherent, AFE domain patterns, suggesting a stabilizing effect on the AFE phase. This is entirely consistent with expectations based on the enhanced AFE character of macroscopic *P-E* hysteresis loop measurements. The recent realization of well-defined double hysteresis loops in NN-*x*SS solid solutions has demonstrated the role of Sn$^{2+}$ and Sr$^{4+}$ substitution in stabilizing the AFE order, similar to PZLST solid-solutions, achieving a remarkable 8-times enhancement of the recoverable energy density.[4,9] Furthermore, *Ding et al.* also observed that the domain morphology in the NN-*x*SS series transforms from microsized domain blocks in NN to "parallelogram domains" delineated by sharp domain/twin boundaries in NN-0.05SS.[4] As such, our own results support the findings of these prior reports.

However, unlike these prior reports, our DFXM data provides further information regarding how these more ordered AFE regions correlate to the local strain environment within the grains of NN and NN-0.05SS. **Figure 2** shows the probability density functions for the elastic strain (i.e. "macrostrain") and FWHM (i.e. "microstrain") of the two AFE compositions, in which clear differences are noticeable.

Although the strain distribution in **Figure 2(a)** indicates that the standard deviation ($\sigma$) of both samples appears to be comparable with $\sigma$ = 0.044, we see that the profile of NN-0.05SS histogram deviates significantly from the unimodal Gaussian distribution of pure NN. Rather, it contains four distinct peaks. Although the overall spread of both samples is apparently within the same interval, the average $\sigma_{NN-0.05SS}$ (*i.e.*, accounting the four peaks) is 0.025 ± 0.008. Therefore, indicating a 43% reduction in the strain heterogeneity at the macroscopic level.

At the same time, the FWHM distributions in **Figure 2(b)** show that SrSnO$_3$ substitution significantly reduces the microstrain, *i.e.* mesoscale heterogeneity. This is because the broadening in FWHM correlates to non-uniform lattice deformations and the distribution of structural defects affecting the long-range structural disorder. Fitting these distributions with lognormal functions reveals that NN-0.05SS reduces the FWHM spread by 41%, while lowering the mean FWHM value by 67%. In short, the formation of $NaNbO_3$-$SrSnO_3$ solid-solution is effective because it promotes more structural ordering at the mesoscale.

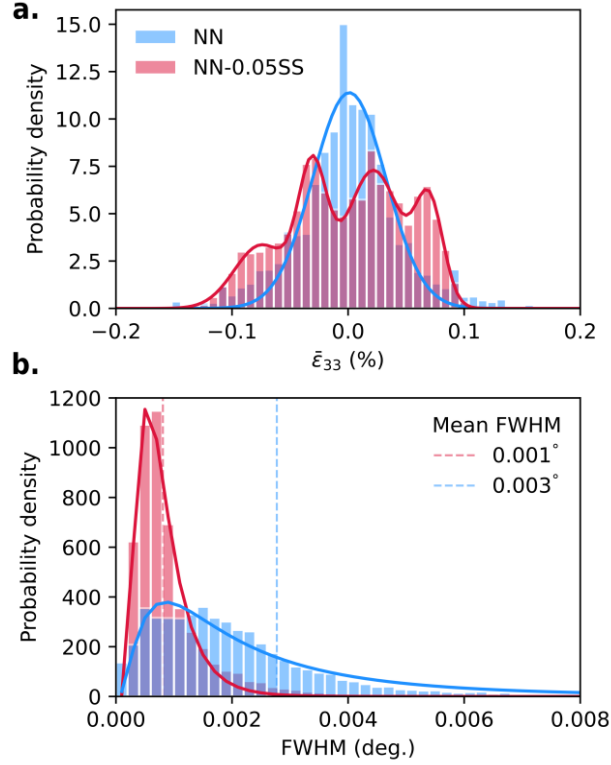

**Figure 2.** Normalized histograms of (a) relative elastic strain ("macrostrain") and (b) FHWM ("microstrain") for NN and NN-0.05SS. The distinct sharper peaks in NN-0.05SS account for a reduced heterogeneity in the macroscopic strain distribution. The solid lines in (a) and (b) correspond to the Gaussian and lognormal distributions, respectively. The reduction in FWHM is significantly more pronounced upon $SrSnO_3$ substitution in contrast to the elastic strain.

Both the FWHM maps in **Figure 1(d)** as well as in other DFXM studies of embedded grains indicate that strain relaxation across boundaries can significantly influence the average degree of structural disorder – particularly for small-grained (< 15 μm) materials.[8] To assess this in our case, **Figure 3** displays the radial mean profile (*i.e.* from pixels positioned equidistantly from the center of mass) of $\bar{\varepsilon}_{33}$ and FWHM from the AFE phase boundary towards its center. It can be seen in **Figure 3(a)** that, despite the strain level of pure NN being well-relaxed at the AFE phase boundaries (with $\bar{\varepsilon}_{33}$ close to zero), the strain tends to fluctuate around positive values towards the center. Such elevated strains in the grain interior are likely a consequence due to the internal interfaces between the major regions (likely ferroelastic domains). On the other hand, NN-0.05SS shows a much larger variation in $\bar{\varepsilon}_{33}$ strains across the AFE phase boundary, and only appears to plateau very close to the center at a compressive strain of -0.076%.

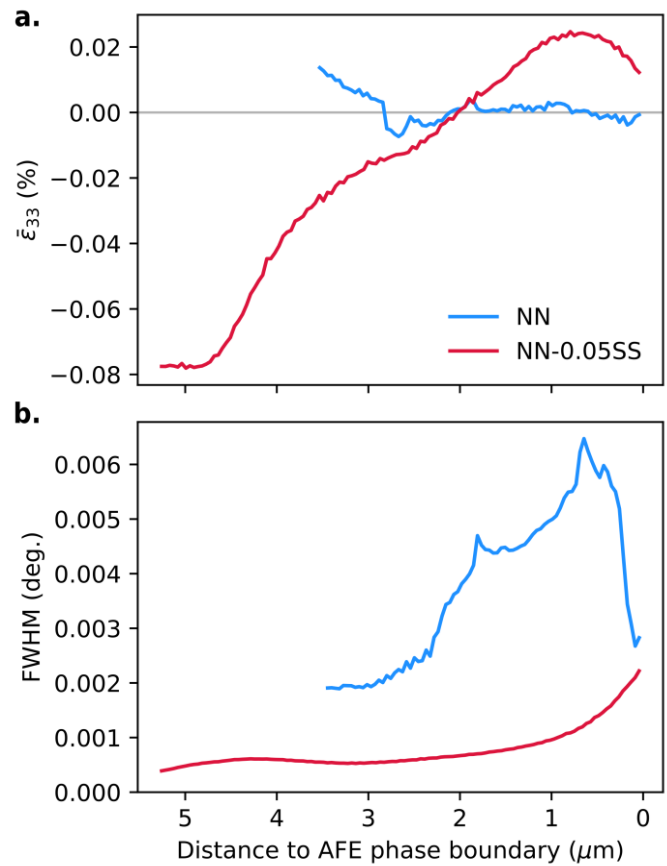

**Figure 3.** Mesostructural relaxation as the radial mean profile of (a) $\bar{\varepsilon}_{33}$ and (b) FWHM, extending from the AFE phase boundary towards their center of mass.

In contrast, the radially-averaged FWHM in **Figure 3(b)**, which indicates how the relaxation due to defects (*i.e.* the main origin of FWHM broadening) behaves within the mesoscale, depicts the opposite trend. Although pure NN exhibits a large variation in microstrain/disorder, the relative variation of FWHM between the AFE phase boundary and its center is minimal. These factors indicate a negative impact on the overall AFE phase stability. On the other hand, the relaxation profile of NN-0.05SS is closer to what one would expect for a typical ceramic grain, where the structural disorder primarily originates from the structurally incoherent interfaces at their boundaries.

One should exhibit caution when deriving conclusions from radially-averaged measurements taken from only a few grains with different sizes and geometries. However, the results here are strongly consistent with the picture that 5 mol.% of $SrSnO_3$ substitution serves to stabilize the structure such that it can maintain AFE order across a broader range of residual stresses caused by defects such as grain/phase boundaries and twin walls.

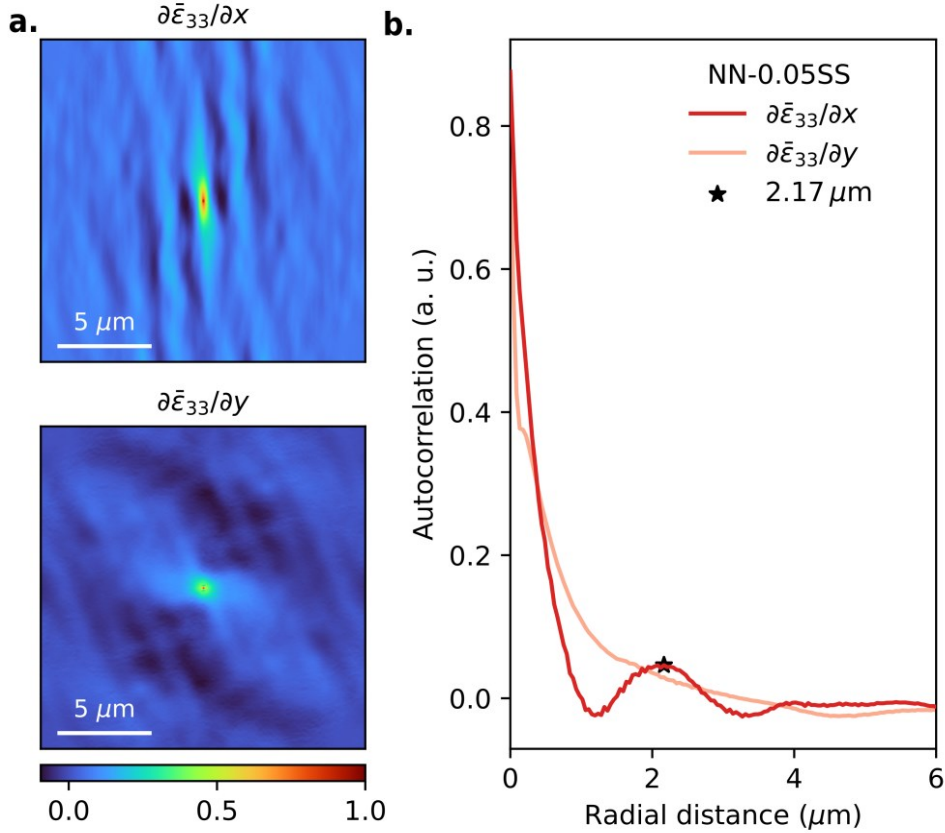

**Figure 4.** (a) The sinusoidal fluctuations in the autocorrelation maps of $\partial\bar{\varepsilon}_{33}/\partial x$ and $\partial\bar{\varepsilon}_{33}/\partial y$ for NN-0.05SS. (b) Radial average autocorrelation profile of the maps shown in (a).

The increased degree of chemical disorder as a result of substitution with 0.05SS also facilitates an increase in the average domain size to the extent that the domains are directly visible via DFXM (shown in **Figure 1(c)**). These domains are remarkably consistent in size and shape, and can be quantified from the strain gradient maps, $\partial\bar{\varepsilon}_{33}/\partial x$ and $\partial\bar{\varepsilon}_{33}/\partial y$, by computing their radially averaged autocorrelations maps from two-dimensional Fourier Transformations, shown in **Figure 4(a)**.[16,17] Both gradient components display evidence of periodicity, however, the periodicity in $x$ is more prominent due to the asymmetry spatial resolution function of our microscope (we expect better resolution in $x$ than in $y$ or $z$). Ultimately, **Figure 4(b)** indicates a pseudo-period of 2.17 μm and a characteristic feature size (*i.e.,* the position of the first valley) of 1.04 μm. This characteristic feature size is close to the resolution limit of the DFXM experiment and is in accordance with the size of the "parallelogram" domains found by *Ding et al.*.[5] Similar observations could not be found for NN (not shown here) due to the high heterogeneity, however one would expect AFE domains to be smaller and with lower periodicity.

In conclusion, our results reveal two significant findings: firstly, the ability of DFXM to utilize superstructure (or satellite) reflections to map strain heterogeneity in complex bulk granular mesostructures that would normally be invisible when imaging with fundamental reflections. Moreover, it would allow the direct visualization of the order parameter, given that superstructure reflections are signatures of vibrational modes and phase transitions. Secondly, our novel approach clearly shows how lead-free antiferroelectric NN solid solution with 0.05SS can dramatically enhance the degree of mesostructural order and promote the formation of stable AFE domains with clear periodicity in the strain gradient. In the case of pure NN, we find that the surprisingly heterogeneous microstructure restricts the proper nucleation of AFE domains/phase, forcing the microstructure to accommodate an uneven distribution of defects within domains, resulting in a poor structural relaxation that would ultimately compromise the ability to exhibit reversible AFE-FE phase transitions and the characteristic double hysteresis loops of an effective AFE-*based* energy storage material.

We view these results as an encouraging outcome for both the engineering of new and the improvement of existing lead-free AFE materials, whether based on binary, ternary, or even high entropy solid solutions. Additionally, these outcomes are promising for the characterization of materials presenting specific functionalities associated to the control of specific modulation periods and lengths.

**Acknowledgments**

The authors are grateful to ESRF for providing beamtime at ID06-HXM under proposal MA-4442. L.O. and H.S. acknowledge financial support from ERC Starting Grant #804665. C.Y. acknowledges financial support from ERC Starting Grant #101116911. This work was partially supported by the Hessian State Ministry for Higher Education, Research, and the Arts under the LOEWE collaborative project FLAME (Fermi level engineering of antiferroelectric materials for energy storage and insulation systems).